\documentclass{aastex631}

\shorttitle{The $M_{*}-M_{\rm halo}$ relation at 0.08 $<z <$ 1.53 in COSMOS}
\shortauthors{Vardoulaki et al.}

\graphicspath{{./}{figures/}}

\begin{document}

\title{The $M_{*}-M_{\rm halo}$ relation at 0.08 $< z <$ 1.53 in COSMOS: the role of AGN radio-mode feedback}

\correspondingauthor{Eleni Vardoulaki}
\email{elenivard@gmail.com}

\author[0000-0002-4437-1773]{Eleni Vardoulaki}
\affiliation{Th\"{u}ringer Landessternwarte, Sternwarte 5, 07778, Tautenburg, Germany}

\author[0000-0002-0236-919X]{Ghassem Gozaliasl}
\affiliation{Department of Physics, University of Helsinki, P. O. Box 64, FI-00014, Helsinki, Finland
}

\author[0000-0002-4606-5403]{Alexis Finoguenov}
\affiliation{Department of Physics, University of Helsinki, P. O. Box 64, FI-00014, Helsinki, Finland
}

\author[0000-0002-2640-5917]{Eric F. Jim\'{e}nez-Andrade}
\affiliation{ National Radio Astronomy Observatory, 520 Edgemont Road, VA 22903, Charlottesville, USA
}

\collaboration{6}{and the COSMOS Team}

\begin{abstract}

In the current picture of cosmology and astrophysics, the formation and evolution of galaxies is closely linked to that of their dark matter haloes. The best representation of this galaxy-dark matter halo co-evolution is the $M_{*}-M_{\rm halo}$ relation. In this study we investigate how the radio-mode feedback from active galactic nuclei (AGN) affects the $M_{*}-M_{\rm halo}$ relation  at redshifts 0.08 $< z <$ 1.53. We use a set of 111 radio-selected AGN at 3 GHz VLA-COSMOS within the X-ray galaxy groups in the COSMOS field. We compare these results to the ones of 171  star-forming galaxies (SFGs), using the theoretical relation of \cite{moster13}. We  find that AGN agree within 1\% with the \cite{moster13} relation, SFGs show an offset of 37\%, suggesting that the radio-mode feedback from AGN at a median redshift of $\sim$ 0.5 still plays a significant role in the $M_{*}-M_{\rm halo}$ relation.

\end{abstract}


\section{Introduction} \label{sec:intro}

In our current understanding of galaxy formation and evolution, in a lambda cold dark matter ($\Lambda$CDM) Universe, galaxies are born within dark matter haloes and co-evolve with their haloes. The best representation to study both dark matter and baryonic matter and probe the underlying physics is through the $M_{*}-M_{\rm halo}$ relation \citep[e.g.][]{behroozi19, moster18}. The $M_{*}-M_{\rm halo}$ relation  reveals the physics behind the coupling of dark matter to baryonic matter, the physics of galaxy formation through physical properties (e.g. stellar mass, star-formation rate SFR, halo mass), as well as how efficiently a galaxy converts baryonic matter into stars. The latter is known as star-formation efficiency $f_{*} = M_{*}/f_{\rm b}M_{\rm halo}$, and is strongly linked to the dark matter halo. 

The star-formation efficiency peaks for halo masses around $10^{12} M_{\odot}$ (Milky-Way-type galaxies), while the maximum efficiency galaxies can reach is rather small, near 20\%. This implies that almost most of the baryons are undetected even in the systems that most efficiently convert baryons to stars. The small $f_{*}$ values are attributed to feedback mechanisms, which play an important role in galaxy growth and evolution. In order to avoid having overly massive galaxies in the local universe, star formation needs to be regulated via some type of feedback. Radio-mode feedback from active galactic nuclei (AGN) is considered the maintenance mode of feedback \citep{fabian12}. It heats the halo and efficiently quenches star formation (SF), which translates to a small efficiency in which galaxies convert dark to baryonic matter. 

Although models suggest feedback is necessary, recent studies question the role of AGN feedback in the local universe. \cite{posti19} showed that the halo masses associated with massive spirals are smaller than expected from models \citep{moster13} and empirical relations \citep[see][ for a review]{wechsler18}. \cite{posti19} found that massive spirals in the local Universe convert all their baryonic matter in the halo into stars, suggesting that feedback is not as important as previously thought.

This project explores the $M_{*}-M_{\rm halo}$ relation at a median redshift of 0.5 in the Cosmic Evolution Survey (COSMOS) field, in order to take advantage of the plethora of ancillary multi-wavelength data. A low-density, $\Lambda$-dominated Universe in which $H_{0}=70~ {\rm km~s^{-1}Mpc^{-1}}$, $\Omega_{\rm M}=0.3$ and $\Omega_{\Lambda}=0.7$ is assumed throughout.

\section{Sample}

To perform our analysis we cross-correlate the radio sources from the VLA-COSMOS survey at 3GHz \citep{smolcic17, vardoulaki19, vardoulaki21} with the X-ray galaxy group catalogue of \cite{gozaliasl19}. The X-ray Chandra observations probed 247 X-ray galaxy groups in the redshift range 0.08 $< z <$ 1.53, and halo masses for the groups in the range $M_{\rm 200c} = 8\times10^{12} - 3\times10^{14} M_{\odot}$ ($M_{\rm 200c}$ is the mass of the group within the virial radius $R_{200}$). Extended radio sources are classified based on the FR-type classification scheme \citep{fanaroffriley74} taking into account the surface brightness distribution, in FRI, FRII, and FRI-FRII \citep{vardoulaki21}, and jet-less objects are separated in compact AGN (COM AGN) or SFGs based on their excess radio emission compared to what is coming from star-formation alone \citep{delvecchio17, vardoulaki21}. The cross-correlation yields 24 (out of 75) FRs, 87 (out of 963) COM AGN, and 171 (out of 6452) SFGs. Host stellar mass was calculated from fitting the infrared spectral energy distributions \citep{smolcic17b}. The halo mass of each host is assumed to be the X-ray group halo mass in the case the host is the brightest group galaxy (BGG). Otherwise the host halo mass is an upper limit. The fraction of radio galaxies at the high stellar-mass end ($M_{*} > 10^{10.5 }M_{\odot}$) is $\sim$20\%, and drops below $\sim$9\% for small stellar masses. We note that the sample is highly affected by incompleteness below stellar mass of $10^{9} M_{\odot}$.

\begin{figure}[!ht]
\resizebox{\hsize}{!}
{\includegraphics{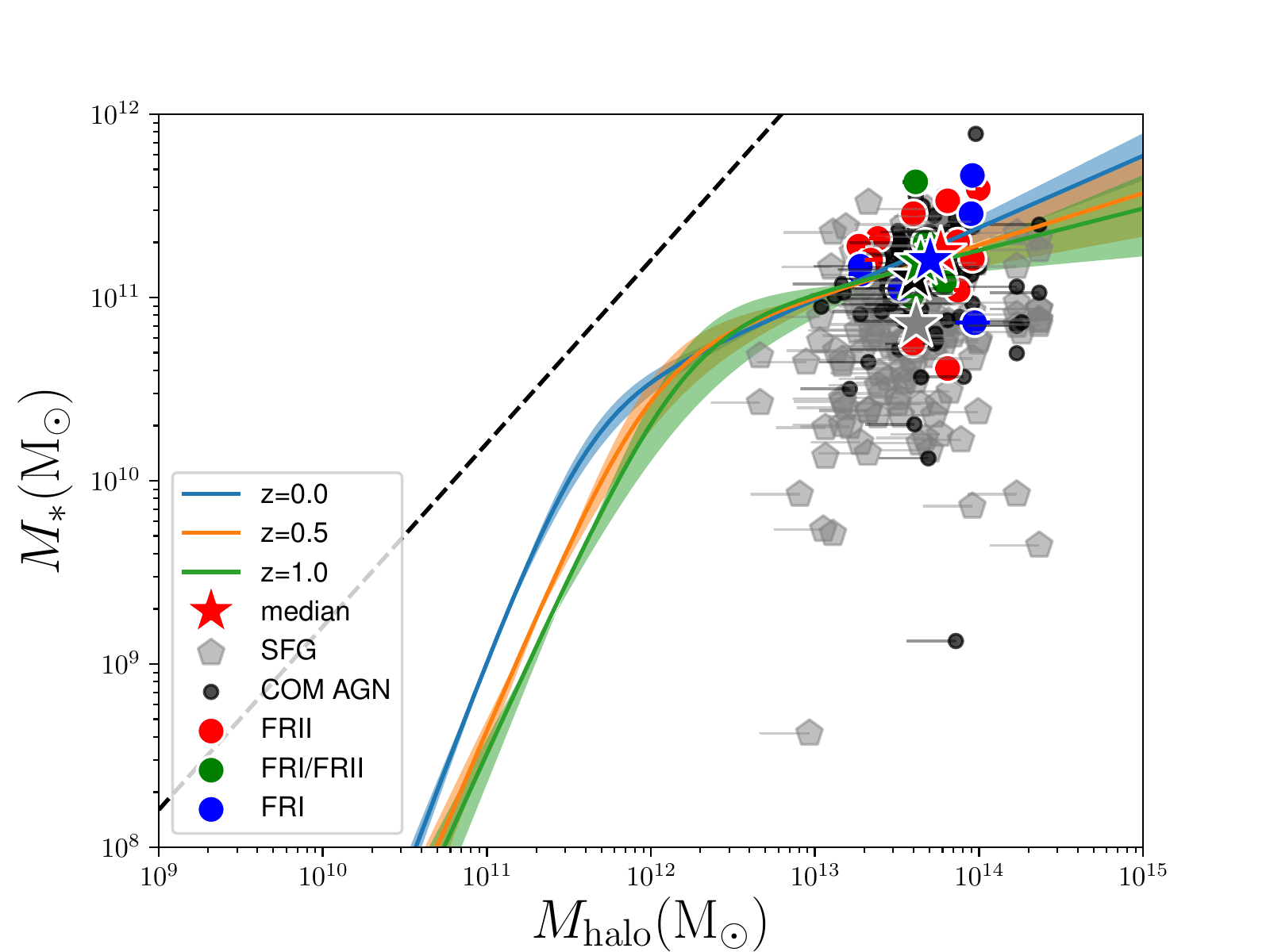}
 \includegraphics{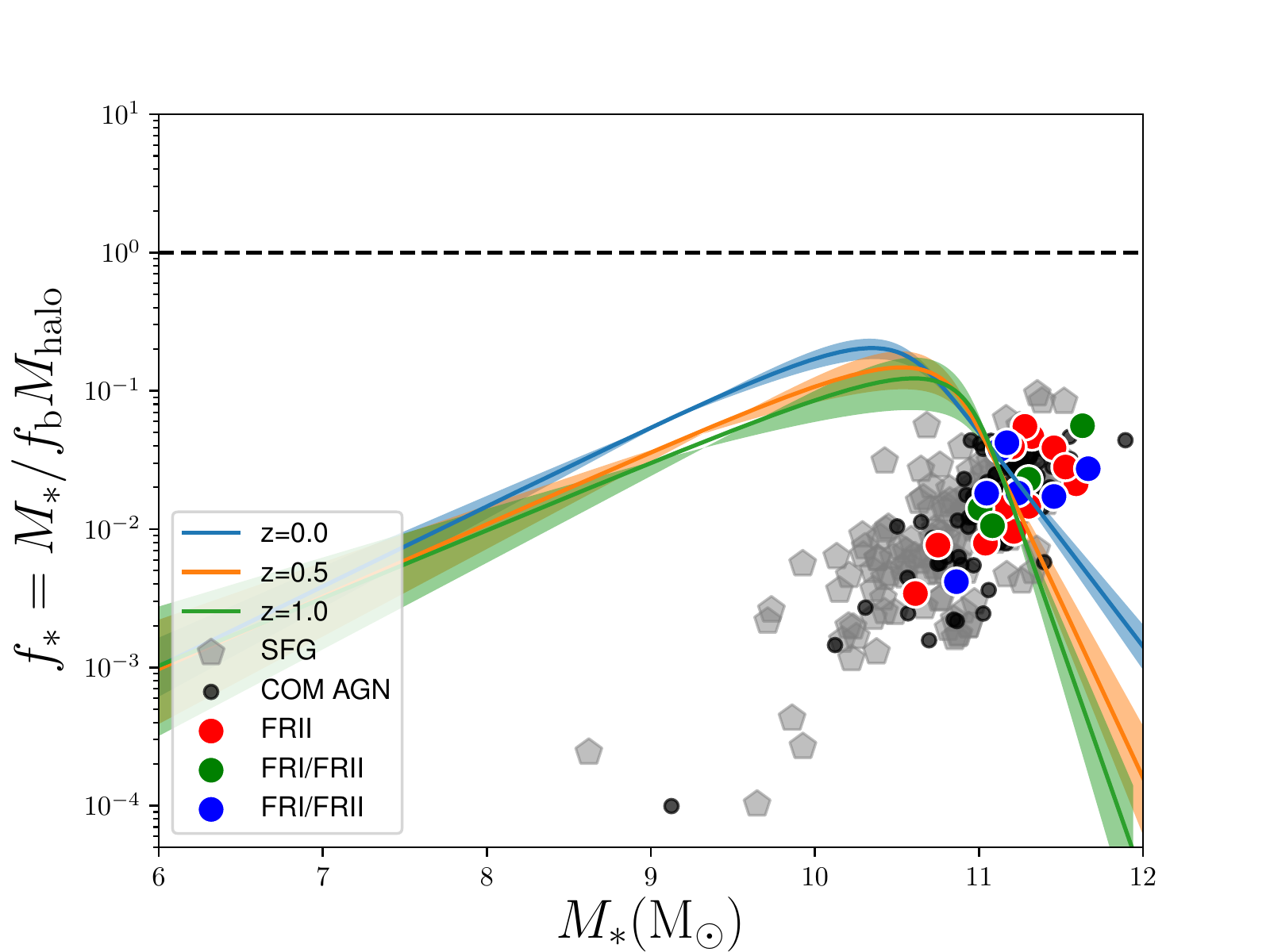}
            }
\caption{{\bf Left:} $M_{*}-M_{\rm halo}$ relation for COSMOS at 0.08 $< z <$ 1.53. Symbols: the SFGs as grey pentagons, the COM AGN as small black circles, and the radio AGN with jets as red (FRII), green (FRI-FRII), or blue (FRI) circles. The dashed line indicates that all baryonic matter in the halo of a galaxy is converted into stars (we assumed $f_{\rm b}$= 0.16). The coloured lines are from the model of \cite{moster13}  for redshifts 0, 0.5 and 1. Limits denote that the host is not a BGG and thus the $M_{\rm halo}$ is an upper limit. Median values are shown as large stars with white edges, where colour corresponds to each sub-sample. {\bf Right:} Star-formation efficiency $f_{*}$ vs stellar mass. The horizontal line is for 100\% efficiency. Symbols as on the left.
   }
\label{fig:mstar}%
\end{figure}

\section{Analysis and Results}

In Fig.~\ref{fig:mstar} we show the $M_{*}-M_{\rm halo}$ relation for the objects in our sample. We compare our sample to the predictions of \cite{moster13} for redshift up to one. We estimate the $M_{*}$ for the median $M_{\rm halo}$ and redshift of each population, based on the \cite{moster13} relation. The median values of our sample are shown in Fig.~\ref{fig:mstar} as stars.  We find a good agreement for FRs and COM AGN with an offset of $<0.1\%$ and 1\% respectively from the \cite{moster13} relation at a median $z = 0.5$, and within the spread of the theoretical relation. For SFGs we find larger scatter, especially for smaller stellar masses ($<10^{10.5}~M_{\odot}$), and a larger offset of 37\% from the \cite{moster13} relation at a median $z = 0.5$. 

We are probing a different parameter space than local studies \citep[e.g.][]{posti19}, which showed there is no AGN feedback needed to regulate star formation. Our results suggest that feedback is in place, and that the objects in our sample follow the theoretical models up to redshift of 1, albeit with a large scatter. This scatter can be attributed to the $M_{\rm halo}$ being the halo mass of the associated X-ray galaxy group, and thus an upper limit. The most massive galaxies, associated with AGN, follow well the predictions, while SFGs display large scatter, especially at low stellar masses.

\section{Discussion and Conclusions}

In this project we investigated the $M_{*}-M_{\rm halo}$ relation at 0.08 $< z <$ 1.53 with COSMOS radio \citep{vardoulaki21} and X-ray \citep{gozaliasl19} observations. We separate objects in radio AGN with jets (FR), radio AGN without jets (COM) and star forming galaxies (SFG). We plot the $M_{*}-M_{\rm halo}$ relation for the objects in our sample, and we find agreement with predictions \citep{moster13}, with a large scatter which is attributed to the $M_{\rm halo}$ values being the ones of the associated X-ray group. In \cite{vardoulaki21} we found indications of radio-mode feedback related to the FR-type objects in our sample occupying massive hosts ($> 10^{10.5}~M_{\odot}$), where objects with the earliest episodes of star formation are the ones leaving the main sequence for SFGs. 

Our results at $z_{\rm med} \sim 0.5$ using COSMOS data suggest that SF efficiency decreases at the high-mass end, which agrees with theoretical expectations, and further supports the role of AGN feedback in regulating the growth of massive galaxies. Future deeper surveys can probe the Milky-Way-type galaxies beyond $z \sim$ 0.5 and also probe the peak of the $M_{*}-M_{\rm halo}$ relation, in order to better compare to local studies and further investigate the role of AGN feedback on the $M_{*}-M_{\rm halo}$ relation.


\bibliography{mstarmhalo}{}
\bibliographystyle{aasjournal}



\end{document}